# RESEARCH PAPERS

# THE ROLE OF CLIENT ISOLATION IN PROTECTING WI-FI USERS FROM ARP SPOOFING ATTACKS

By

TIMUR MIRZOEV *                                            JOEL STACEY WHITE **

*Professor of Information Technology Department, Georgia, Southern University, United States.*
*\*\* Office of Information & Instructional Technology at Bainbridge College.*

## ABSTRACT

*This study investigates the role of the client isolation technology Public Secure Packet Forwarding (PSPF) in defending 802.11 wireless (Wi-Fi) clients, connected to a public wireless access point, from Address Resolution Protocol (ARP) cache poisoning attacks, or ARP spoofing. Exploitation of wireless attack vectors such as these have been on the rise and some have made national and international news. Although client isolation technologies are common place in most wireless access points, they are rarely enabled by default. Since an average user generally has a limited understanding of IP networking concepts, it is rarely enabled during access point configurations. Isolating wireless clients from one another on unencrypted wireless networks is a simple and potentially effective way of protection. The purpose of this research is to determine if a commonly available and easily implementable wireless client isolation security technology, such as PSPF, is an effective method for defending wireless clients against attacks.*

*Keywords: Wireless Networking, Wi-Fi, Information Security, ARP Poisoning, PSPF.*

## INTRODUCTION

The use of Wireless Networking Technologies to communicate on the Internet continues to rise every year, with public hotspots numbering in the tens of millions worldwide (WeFi., 2011). Laptop computers and smartphones using the 802.11 (Wi-Fi) protocols make up a large number of the millions of wirelessly connected devices (Kendrick, 2010). Wireless networking is a convenient way to stay connected from remote and mobile locations (Bowman, 2003). The usage of Wireless Local Area Networks (WLAN) is widely spread due to the fact that computers communicate wirelessly, without the need for physical cable connectivity (Umar, 1993). Unfortunately, since the data is transferred over public airwaves where it can theoretically be intercepted by others within range, wireless networking creates a unique set of security concerns.

Connections between Wi-Fi enabled clients and network access points (APs) can be secured through a variety of different methods (Ciampa, 2006). One method, Wi-Fi Protected Access (WPA), involves using strong data encryption and rotating authentication keys (Rittinghouse, Ransome, 2004). WPA is very commonly supported by most AP hardware. In the WPA scenario, a wireless client is provided with a temporary encryption key or 'session key' with which it can communicate securely with the AP and encrypt/decrypt traffic (Cayirci, Rong, 2009). Once a time limit is reached or the session has ended the key is expired or exchanged for a new one, making any further use of the key invalid (Cayirci, Rong, 2009). This type of protection prevents an attacker from viewing traffic or capturing a client's key and then using it to impersonate the client. There are other types of security of this nature, but they all have one thing in common: it is necessary for the client to authenticate to the AP in some way with pre-shared authentication information.

This study determines if one manufacturer's version of client isolation, Cisco System's Public Secure Packet Forwarding (Cisco, 2010), prevents wireless devices from launching impersonation attacks. The focus of this study is on the well-known man-in-the-middle IP ARP spoofing attack (Cross, 2008). It is generated using a publically available set of hacking tools known as Cain & Abel (Oxid, 2011). Since Cisco's Aironet access points are widely used throughout the world, and Cain & Abel is free to download from the Internet, such a combination seems to constitute a likely real-world scenario and test of the premise of client isolation protection.

### 1. Literature Review

In the case of clients using publically available Wi-Fi connections, authentication is not only inconvenient but is impossible due to a large number of unique visitors. The convenience of attaching to a public Wi-Fi network at will without hassle is what makes the technology inviting to so many people. However, there is a major problem with this type of scenario: Public Wi-Fi connections without authentication transmit and receive traffic in an unencrypted or plain-text fashion. This means that anyone listening with a Wi-Fi data capture device can intercept the traffic and read its contents. This is not the only issue involved. Well known vulnerabilities could allow an attacker to place himself/herself in the middle of a user's conversation, in effect 'becoming' the user on the wireless network (Ali, 2008). There are ways in which this type of scenario can be avoided: Through the use of a technology called 'client isolation'. Client isolation works on the premise that any wireless device connected to a Wi-Fi access point must communicate through the AP in order to talk to any other device, whether that other device is located on the local area network (LAN) to which the AP is connected or on the Internet. Using this fact to their advantage, some Wi-Fi AP manufacturers make it possible to isolate clients from each other by preventing any device associated to the AP from talking to any other associated device, using the AP transceiver as the moderator. This does not prevent Wi-Fi-based attacks on the LAN, or LAN data sniffing, but it does in theory prevent wireless devices on a particular AP from directly attacking and impersonating each other.

In recent years, substantial negatives associated with the use of wireless communication technologies have come to light. Specifically, a simple social networking account hacking tool known as Firesheep, released in October of 2010, has made international news (Gahran, 2010). Firesheep is a freely downloadable extension to the FireFox web browser developed by Eric Butler (Butler, 2010). Firesheep allows its users to capture other wireless users HTTP session cookies, and then access that captured person's account information. Authentication attack with Firesheep scenario has been previously presented by Miller and Gregg (Miller, Gregg, 2011). Although Firesheep is designed to allow a quick access to a predefined number of social networking sites, its open source nature also allows it to be modified for other uses. This tool uses a hacking technology known as 'side-jacking' - it does not hijack an HTTP session; it only captures the session's cookie to allow forging of a new 'side' connection (Leonhard, 2010). Many popular social networking sites, such as Facebook and Twitter, do not encrypt their communications by default, making this sort of attack possible (Vaughan-Nichols, 2010).

There is another form of attack that is not well known to an average wireless user. It is far more dangerous and disturbing than ones such as Firesheep from a security perspective. That attack is known as an ARP spoofing or ARP cache poisoning attack. It is a classic man-in-the-middle session hijacking, which can also be executed through freely downloadable software such as Oxid's Cain & Abel (Sanders, 2010). ARP spoofing allows a malicious user on the network to place their computer's network interface between a victim and the computer(s) with which they are communicating (Graves, 2007). This insertion allows for a capture, and possible modification, of all data sent; in essence completely hijacking the conversation. A hacker can then pose as the intended recipient, corrupt the data sent, change that data, or simply just capture it for later use (Marcus, 2010). The main difference between this attack and the Firesheep attack is that the hijacked users have not just inadvertently allowed access to their social networking account; they have placed themselves at the mercy of the attacker, possibly giving up credit card numbers and/or private information, completely unaware of the situation.

Since a Firesheep attack and ARP spoofing require open and unsecured wired or wireless communications to operate effectively, there is a need to find proper protection against such occurrences. The Firesheep problem is solvable by the web sites that the users are accessing, and some have already taken steps to make the required fixes (Marcus, 2010). Encrypting all communications between users and servers through the use of web session encryption technologies like SSL (Secured Sockets Layer) eliminates the usable capture of session cookies in transit, thereby protecting from Firesheep. ARP spoofing attacks on an open access wireless network are much different in nature (Cross, 2008). Since an attacker can intercept users' data, it is possible for even encrypted communications to be hijacked (Sanders, 2010). In such a scenario a different type of security technology is needed.

Cisco Systems, a world-wide leader in the manufacture of networking equipment (Cisco, 2011), has implemented such a technology in most of its wireless access point hardware; that technology is known as PSPF, or Public Secure Packet Forwarding. Another non-Cisco term for PSPF exists and it is called Intra-BSS (Putman, 2005). PSPF is a client isolation technology intended to prevent one wireless client from talking to another. This technique is used to prevent direct client-on-client attacks, but should also prevent session hijacking by stopping spoofed ARP packets from reaching the victim. Since Cisco technology is very commonly deployed by small to very large businesses, it seems like a logical choice to base this research upon. It is also possible, through the use of third party software products such as ArpOn (Di Pasquale, 2011), for clients to protect themselves from ARP spoofing attacks. However, many public wireless users are either not skilled or informed enough to take such precautions. An effort on the part of the network owners to secure the communications of their users makes more sense, from a security and efficiency standpoint. Implementing the protective solution inside of the access point hardware, instead of relying upon the users to do it, will secure all users connecting to it, regardless of their configurations.

## 2. Methodology and Setup

To demonstrate the ability of PSPF to protect Wi-Fi clients from ARP spoofing attacks, a test attack was simulated using two Dell Wi-Fi enabled laptops and a Cisco access point. The Cisco access point used in this study was a Cisco Aironet model 1232-AG. The access point was restored to the factory default configuration, and was configured to provide two open access Wi-Fi connections to public clients: 'FREE_WIFI', which was unsecured by PSPF, and 'FREE_WIFI_PSPF', which was secured by PSPF. These two Wi-Fi connections were accessed by two Dell laptop clients, a Dell Precision M4500 which acted as the victim, and a Dell Latitude D420 which acted as the attacker. Both Dell laptops were configured with the Microsoft Windows 7 operating system, the latest operating system patches, and 802.11G Wi-Fi network adapters. The attacking computer was also configured with Wireshark, a data packet sniffing freeware product (Wireshark, 2011), and Cain & Abel, a freeware ARP spoofing hacking tool (Oxid, 2011). These and other details are included in Table 1.

The computers used in this study were chosen because they are typical computers which might be used by public Wi-Fi clients. Dell computers are very commonly used by businesses around the world. Windows 7, the operating system on these computers was chosen for its commonality on most computers sold today - Microsoft Windows is the market leader for operating systems, holding an almost 90% market share (Brodkin, 2011).

The freeware tools were chosen based on their popularity, usability, and availability. Wireshark is a widely known and used Ethernet packet capture and analysis tool, once named Ethereal, and is freely available to the public for download (Wireshark, 2011). Cain & Abel is a well-known and commonly used hacking tool suite, which is also available for download free of charge (Oxid, 2011). The freeware tools were chosen based on their popularity, usability, and availability. Wireshark is a widely known and used Ethernet packet capture and analysis tool (Orebaugh, 2007). Once named Ethereal, it is freely available to the public for download (Brodkin, 2011). Cain

| | Victim's Computer | Attacker's Computer | Wi-Fi Access Point |
|---|---|---|---|
| Manufacturer: | Dell | Dell | Cisco |
| System Model: | Precision M4500 | Latitude D420 | Aironet 1232-AG |
| Processor Type: | Intel i7 | Intel Core Duo | PPC 405GP |
| Memory Installed: | 8GB RAM | 2GB RAM | 32K NVRAM |
| Operating System: | Windows 7 Ultimate | Windows 7 Pro | IOS 12.3(8)JA2 |
| Network Adapter: | Intel Centrino 6250 | Dell 1490 Dual Band | 802.11-G Radio |
| Web Browser: | Internet Explorer 8 | Internet Explorer 8 | N/A |
| IP Address: | 192.168.13.91 | 192.168.13.92 | 192.168.13.1 |
| MAC Address: | 00-23-15-9B-02-38 | 00-16-CF-B1-51-B1 | 00-1B-54-AB-04-3B |
| Firewall: | Enabled | Disabled | N/A |
| Modifications: | None | Installed Wireshark | None |
| | | Installed Cain & Abel | |

Table 1. Utilized Hardware and Setup

& Abel is a well-known and commonly used hacking tool suite (Wallingford, 2006). It is also available for download free of charge (Oxid, 2011).

3. Testing Procedure

The test began by configuring the Cisco access point, as described above, with two connections, one with PSPF and one without. Both, the attacker and victim computers were connected to the AP using the 'FREE_WIFI' non-PSPF connection. The attacking computer initiated a Wireshark packet capture on the network, which showed ARP broadcast data packets coming from the address of the victim (Figure 1).

Knowing that the client was connected, the attacking computer then launched Cain & Abel, and a network scan was performed. There is a Media Access Control (MAC) address assigned to every network card. It is a unique address by all vendors for every Network Interface Card (NIC). The IP and the MAC address of the victim's computer was discovered, as well as the MAC address of the subnet gateway (Figure 2).

Once this information was logged in Cain & Abel, ARP cache poisoning has begun, making the client use the attacker as its network gateway. Internet Explorer was launched on the victim's computer, connecting to the Google website. The attacking computer showed instantly that web traffic from the victim had been captured. The victim's computer then moved from Google to Gmail, where it logged into a test email account, and read a message; the SSL encrypted email username and password had been captured by the attacker (Figure 3).

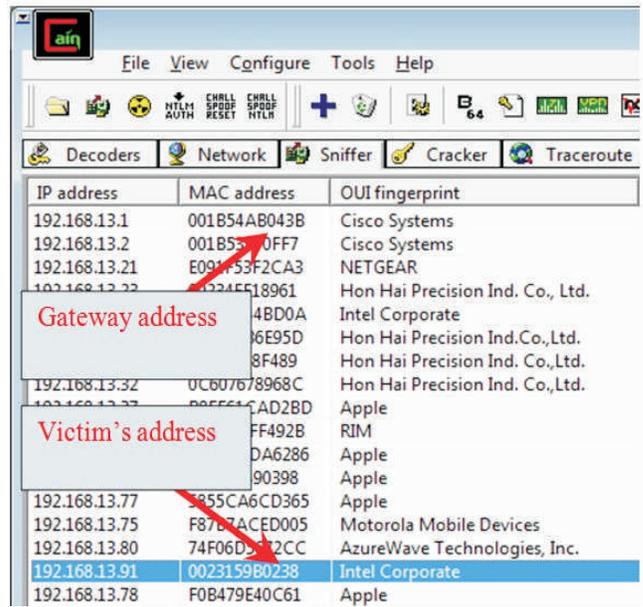

Figure 2. Network MAC Scan without PSPF

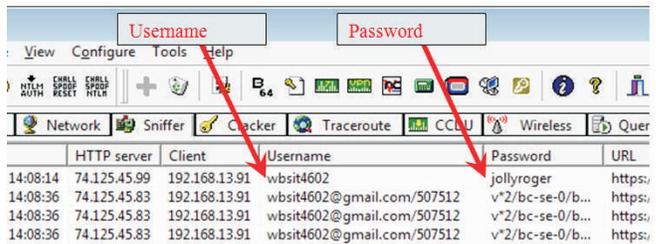

Figure 3. Credentials for Gmail Account Captured

The start of the APR table poisoning can be seen on the Figure 4.

Cain & Abel by default is configured to capture and log HTTP account credentials and to insert false certificates into SSL communications in order to capture and relay that traffic. This insertion of false certificates will cause certificate warnings on most browsers, since the root certificate authority of the false certificate will more than likely not be trusted by the victim computer. The certificate presented by the attacker in this case showed as being valid but untrusted; to simulate a user unfamiliar with such technology, the warning was ignored. Figure 5 represents the insertion of the false certificates.

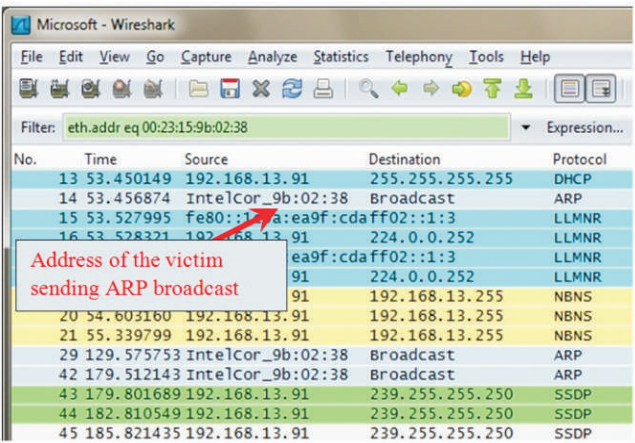

Figure 1. Packet Capture without PSPF

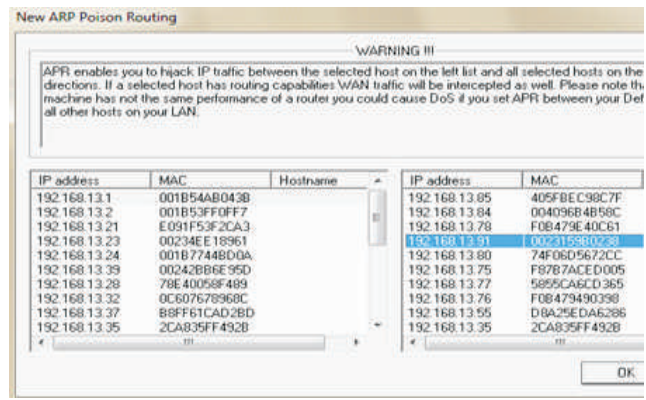

Figure 4. Start ARP Table Poisoning

Later, the test computers were attached to 'FREE_WIFI_PSPF', a PSPF secured wireless connection. The Wireshark packet capture was launched again, but produced no results from the victim computer (see Figure 6). Several packets were seen from the gateway, but none of them were from the victim's computer, not even a broadcast packet.

Cain & Able was then launched and a network scan was again performed. The victim's computer, again, did not show up in the scan results. Since the IP and MAC of the victim were already logged in Cain & Abel from the previous attack, an ARP spoof was attempted using that information. The ARP spoofing failed completely (Figure 7), producing only a brief denial of service for the victim computer. Eventually the victim computer and the network gateway resolved this error, allowing the web session to continue unhindered. No data packets intended for the victim's computer were captured by the attacker during the ARP spoofing phase of the test, and the victim remained invisible to the attacker's network scans and packet sniffing.

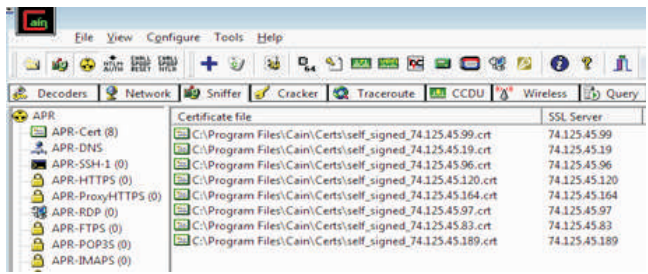

Figure 5. Example of false certificates injected by attacker

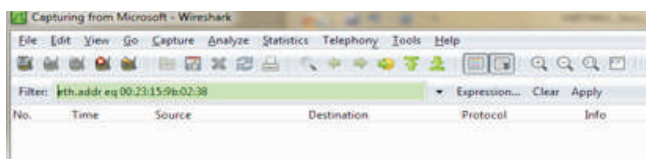

Figure 6. Wireshark capture with PSPF enabled produces no packets from victim

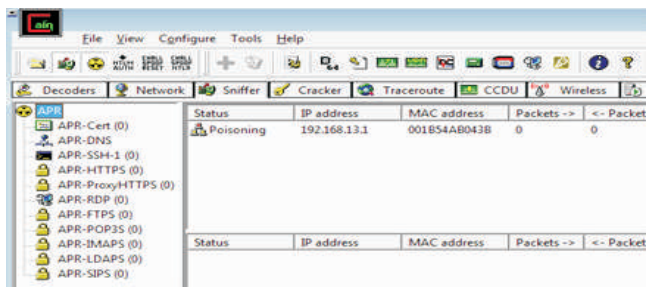

Figure 7. ARP Cache Poisoning with PSPF enabled fails to poison client routes

Conclusion

The tests performed in this study show the role of a client isolation technology, such as PSPF, in protecting 802.11 Wi-Fi clients from ARP spoofing attacks. ARP spoofing attacks are dangerous in that they are difficult to detect, allow an attacker unrestricted access to a victim's network communications, and can be easily executed over public unsecured wireless connections. Since many people around the world use publicly available wireless connections to access the Internet, an attack of this sort could put their personal and confidential information at risk if those connections are not secure.

It was hypothesized that by implementing a client isolation security technology on the wireless access point, and thereby preventing direct client-to-client communication, such attacks would be rendered useless. Since ARP spoofing attacks require the attacker to send forged ARP-reply information directly to the victim in order to corrupt its ARP cache, the attack would not work because the victim would never receive that information. As an additional benefit, the attacker would not be able to send data of any kind to other clients, preventing the attacker from scanning the access point for potential victims and probing their service ports. Further, the victim would also be unable to send data of any kind to the attacker, preventing the attacker from sniffing potential victims' broadcasts from the network. The attacker would be effectively isolated on the access point, unable to see other wireless users or their network traffic.

Without client isolation the attacker easily saw the broadcasts of the potential victim, corrupted or 'poisoned' its ARP cache, and misled the victim into relaying its network traffic through the attacker's computer. With the client isolation technology PSPF enabled on the access point, even though the attacker could not see potential victims, it was still unable to corrupt the ARP cache of a victim that was known to be active on the network. This suggests that in this case, PSPF did in fact prevent an ARP spoofing attack from taking place, and effectively protected the victim from the attacker's advances.

*Recommendations*

Client isolation technologies are available on many different wireless access points, depending on the manufacturer and device complexity. Generally, most high-end enterprise quality access points contain a client isolation option. Unfortunately, many lower-end commodity products do not. As technology becomes more affordable and the thrust for information security becomes more widespread, security features such as client isolation will become more commonplace in the access point market. The feature is generally easy to implement, and can normally be turned on simply by checking the appropriate box under the wireless connection's configuration page on the access point management interface.

Client isolation does have one drawback - it will prevent users of an access point from connecting to wireless peripherals such as wireless printers or storage devices. The solution to this problem is to directly connect these devices to a switch, along with the access point, so that client isolation will not affect them. Since most home or small business wireless access points are connected to a broadband router which contains a switch, this problem can be solved if the need arises.

Despite the fact that the security benefits of wireless access point with PSPF configuration are strong, most manufacturers that provide client isolation leave it turned off by default. This trend needs to change, or a different form of security for public wireless users needs to be developed. In the future, it may be reasonable to design some sort of transport layer security into the wireless access point which allows a public user to have his or her wireless data encrypted without authentication. This may require a major change in the 802.11 wireless technology standard, and switching costs may be too high for most current users. Implementing this type of solution in future Wi-Fi hardware solutions is not beyond the realm of possibility.

For now, client isolation can be used to make public wireless users less vulnerable.

## ABOUT THE AUTHORS

*Timur Mirzoev is currently working as a Professor of Information Technology Department at Georgia Southern University, College of Information Technology. Dr. Mirzoev heads the International VMware IT Academy Center and EMC Academic Alliance at Georgia Southern University and has over 10 years of experience in Information Technology, Administration and Higher Education. Some of Timur's research interests include cloud computing, server and network storage virtualization, disaster recovery. Dr. Mirzoev holds the following certifications:  Vmware Certified Instructor (VCI5,4,3), VMware Certified Professional (VCP5,4,3), EMC Proven Professional, LefthandNetworks (HP), SAN/iQ, A+.*

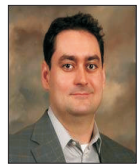

*Stacy White works for the Office of Information & Instructional Technology at Bainbridge College in Bainbridge Georgia. He has been employed there as the Network Administrator / Information Security Officer for the past 12 years. Stacy's Professional interests include Network System Design, Network Security, Risk Assessment, Policy Development, and Industrial Automation and Controls.*

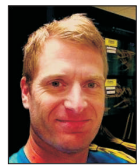